\documentclass{epl}
\usepackage{amsmath}
\usepackage{amssymb}
\usepackage{bm}
\shorttitle{Geometric Phase...}
\title{ Geometric Phase for Fermionic Quasiparticles Scattering by Disgyration in Superfluids}
\author{L.C. Garcia de Andrade
\inst{1}\and A. M. de M. Carvalho\inst{2} \and C. Furtado\inst{2}\thanks{E-mail: \email{furtado@fisica.ufpb.br}} }
\institute{
  \inst{1} Departamento de F\'{\i}sica Te\'{o}rica
            Instituto de F\'{\i}sica-UERJ\\
Rua S\~{a}o Fco. Xavier 524, Rio de Janeiro, RJ,
Maracan\~{a}, 20550-003 , Brazil.\\
  \inst{2} Departamento de F\'{\i}sica,
           Universidade Federal da Para\'{\i}ba\\
           Caixa postal 5008, 58051-970,
            Jo\~ao Pessoa,PB, Brazil\\
  }
\begin{document}
\maketitle
\begin{abstract}
We consider a Volovik's analog model for description of a topological defects in a superfluid and we investigate the scattering of quasiparticles in this background. The analog of the gravitational Aharonov-Bohm in this system is found. An analysis of this problem employing loop variables is considered and corroborates for the existence of the Aharonov-Bohm effect in this system. The results presented here may be used to study the Aharonov-Bohm effect in superconductors.
\end{abstract}
Some condensed matter systems that under certain conditions have a behavior similar to a gravitational system are denominated analog models. In recent years a series of analog models were developed.  Unruh~\cite{prl:unruh} proposed a sonic analog of black holes in a classical fluid, the so-called dumb holes. He started from the continuity and the Euler equations for a classical fluid and got a geometric description for the fluid equivalent to the black hole solution. This system has most of the known properties of a black hole, with the advantage that its basic physics is completely known. The most amazing result obtained by Unruh is that a non-relativistic Newtonian fluid propagating in a flat space plus the time is governed by the geometry of a Lorentzian $(3+1)$-dimensional curved space. A variety of physical systems in condensed matter physics have been used as analog    models: Bose-Einstein condensates~\cite{prl:gar}, classical~\cite{prl:unruh,prd:jac} and quantum fluids~\cite{volo,boo}, moving dielectric media~\cite{leon}, non-linear electrodynamics   ~\cite{prd:novello}, etc. In this letter we investigated the quantum dynamics of quasiparticles in the point of view of Volovik's analog model for quantum fluids. Earlier Volovik~\cite{3} investigated a radial disgyration as a cosmic string which would represent a topologically stable linear defect in $^{3}He-A$.  Thus, the radial disgyration provides only the ``gravity" field which acts on the $^{3}He-A$      fermions with the space-time line element given by
\begin{equation}
\label{metric}
ds^{2} = -dt^{2}+\frac{1}{{c_{1}}^{2}}dz^{2}+\frac{1}{{c_{2}}^{2}}[dr^{2}+
\frac{{c_{2}}^{2}}{{c_{1}}^{2}}r^{2}d{\phi}^{2}],
\end{equation}
where $c_{1}$ and $c_{2}$ represent respectively the ``speed of light" orthogonal and parallel to the gap nodes direction in momentum space, here also ($c_{1}<<c_{2}$) in the superfluid. This geometry possess a conical singularity if $c_{1}$ does not coincide with $c_{2}$, this conical singularity is represented by the following curvature tensor 
$\label{curv}
R_{r,\phi}^{r,\phi}=((1-a)/4Ga))\delta_{2}(\vec{r}),$
where $\delta_{2}(\vec{r})$ is the two-dimensional delta function, and $a=c_{2}/c_{1}$. This behavior of curvature tensor is denominated conical singularity~\cite{staro}. The conical singularity gives 
rise to the curvature concentrated on the disgyration axis. This metric as has been pointed out in Volovik's work is similar to the one of a cosmic string. A crucial difference is the fact that, for a disgyration metric, one has $a>1$ and, in the cosmic string case the parameter $a$ is smaller than 1. This is related to the linear mass density $\mu$ by $a=1-4G\mu$. Values of $a>1$ in gravitation are not physically accepted, but in  condensed  matter this ``antigravitating" string can exist in defects in solids~\cite{katanaev} and  in topological defects in superfluid helium. In recent years a geometrical description of defects in solids was presented in the references~\cite{katanaev,1}. In this formalism the defect formation can be viewed as a ``cut and glue" process, known in literature as the Volterra process. The disclination is obtained by either  removing or inserting material in the medium.
In the present case disgyration can be viewed as a defect produced by inserting material in the medium in contrast with a cosmic string that, in this picture, can be produced by removing material. In this way the analog model for disgyration tells us that this defect represents an antigravitating string. 
 
In a metric theory of gravitation, a gravitational field is related to a nonvanishing Riemann curvature tensor. However, the presence of localized curvature can produce effects on the geodesic motion and parallel transport in regions where the curvature vanishes. The best known example of this non-local effect is provided when a particle is transported along a closed curve which encircles an idealized cosmic string~\cite{Vilenkin}. This situation corresponds to the gravitational analogue~\cite{Ford} of the electromagnetic Aharonov-Bohm effect~\cite{pr:aha}. These effects are of non-local origin rather than local and may be viewed as a manifestation of the nontrivial topology of the spacetime of a cosmic string. It is worth calling  attention to the fact that, differently from the electromagnetic Aharonov-Bohm effect, which is essentially a quantum  effect, its gravitational analogue appears also at a purely classical level. Thus, in summary, the gravitational analogue of the electromagnetic Aharonov-Bohm effect, in this context, is the following: particles constrained to move in a 
region where the Riemann curvature tensor vanishes may exhibit a gravitational effect arising from a region of nonzero 
curvature from which they are excluded. 

We also have an analogue of the gravitational Aharonov-Bohm effect when particles are constrained to move in a region where the Riemann curvature does not vanish, but does not depend on certain parameters such as the angular momentum, as in the case a weak gravitational field arising from a rotating massive body (cylindrical shell)~\cite{Frolov} which we are not  considering in this paper. In this situation, we can have gravitational effects associated with this parameter, namely, an analogue of the gravitational Aharonov-Bohm effect. The existence of gravitational Aharonov-Bohm effect in rotating cosmic string was predicted by Mazur~\cite{prl:maz}, he has shown that particles couple to angular momentum density in the string, even though the particle is constrained only to move outside the string core where the Riemannian    curvature vanishes identically. Burgers~\cite{prd:b} and Bezerra~\cite{valdir1} examined the effects of a parallel     transport of vectors and spinors both around a point-like solution and a cylindrically symmetric cosmic string. This procedure gives, in general, non-trivial results. These effects point out to the gravitational analogue Aharonov-Bohm effect. 

The aim of this letter is to investigate, via holonomy transformations, the gravitational~\cite{Ford,valdir1,law} Aharonov-Bohm ~\cite{pr:aha} effect for quasiparticles in the presence of a disgyration in the point of view of Volovik's analog models for superfluids.  The gravitational analog of the Aharonov-Bohm effect was well investigated for quantized vortices in superfluid. This defect play a role of the spinning cosmic string~\cite{prl:maz} and has angular momentum. In the present case the defect does not possesses angular momentum and the Aharonov-Bohm effect arise from the conical singularity in the curvature. 

Now, we employ holonomy transformation to study the  Aharonov-Bohm effect in the background of a disgyration. In the early sixties Mandelstam~\cite{man} proposed a new formalism for electrodynamics  and gravitation in which the fields depend on the paths rather than on space-time points. The fundamental quantity that arises  from this path-dependent approach, the non-integrable phase factor~\cite{wu} (loop variable) represents the field more adequately than the field strength does. In the application of this formalism to the theory of gravity Mandelstam~\cite{man} established several equations involving the loop variables. Some years later Voronov and Makeenko~\cite{vor} showed the equivalence between the equations obtained using this approach and 
Einstein's equations. The quantities we shall consider are the gravitational loop variables and corresponding holonomies of the Christoffel connection or of the spin connection. Keeping the local fields as the fundamental dynamical variables, we shall compute the loop variables in different cases in order to learn about their behavior    and their geometrical meaning. The loop variables in the theory of gravity are matrices representing parallel transport along paths in a space-time with a given affine connection.  When a vector is parallelly propagated along a loop in a manifold $\textsl{M}$, the curvature of the manifold causes the vector, initially at $p\in \textsl{M}$, to appear rotated with respect to its initial orientation in tangent space $T_{p}\textsl{M}$, when it returns to $p$. The holonomy is the path dependent linear transformation $T_{p}\textsl{M}{\rightarrow}T_{p}\textsl{M}$ 
responsible for this rotation. Positive and negative curvature manifolds, respectively, yield deficit or excess angles between initial and final vector orientation under parallel transport around such loops. They are defined as the limit of an ordered product of matrices of infinitesimal parallel transport as
\begin{equation}
U_\nu ^\mu (C_{yx};\Gamma )\equiv \prod\limits_{i=1}^N(\delta _{\rho
_{1-i}}^{\rho _i}-\Gamma _{\lambda _i\rho _{1-i}}^{\rho
_i}(x_i)dx_i^{\lambda _i})  \label{eq1},
\end{equation}
where $x_0=x,$ $\rho _o=\nu ,$ $x_N=y,$ $\rho _N=\mu ,$ $dx_i=(x_i-x_{i-1})/\varepsilon$. The points $x_i$ lie on
 an oriented curve $C_{yx}$ starting at the $x$ and ending at the point $y$. The parallel-transport matrix $U_{\nu ^{}}^\mu$ is a functional of the curve $C_{yx}$ as a geometrical object. The disgyration metric can be written as
$ds^{2}=\eta_{ab}e^{a}\otimes e^{b}$, where the dual 1-form basis are defined by $e^{a}=e^{a}_{\mu}dx^{\mu}$. So, introducing an appropriate basis
\begin{subequations}
\begin{eqnarray}
\label{disgyr-basis}
        e^{0}&=& dt \mbox{,}       \\
        e^{1}&=& \frac{\cos\phi}{c_{2}}dr- \frac{r\sin\phi}{c_{1}}d\phi
\mbox{,}  \\
        e^{2}&=& \frac{\sin\phi}{c_{2}}dr+ \frac{r\cos\phi}{c_{1}}d\phi
\mbox{,}   \\
        e^{3}&=& \frac{dz}{c_{1}}  \mbox{.}
\end{eqnarray}
\end{subequations}
In this way the metric is written as $ds^{2}= (e^{0})^{2}-(e^{1})^{2}-(e^{2})^{2}-(e^{3})^{2}$. We need 
to determine the 1-form connection $\omega^{a}_{b}$ in order to find the holonomy transformations for the "space-time" of a disgyration. The 1-form connection satisfies the Cartan's structure equation,   $T^{a}=de^{a}+\omega^{a}_{b}\wedge e^{b}$,  where $T^{a}$ is the 2-form torsion. Using the fact that the correspondent geometry of  a disgyration is torsion-free, we find the unique non null 1-form connection
\begin{equation}
\omega^{1}_{2}=-\omega^{2}_{1}=\left(1-\frac{c_{2}}{c_{1}}\right)d\phi.
\end{equation}
This result takes us to the following spin connection for the "space-time" of a disgyration
\begin{eqnarray}
\Gamma_{\phi}= \left(
\begin{array}{cccc}
0 & 0 & 0 & 0 \\
0 & 0 & -(\frac{c_{2}}{c_{1}}-1) & 0   \\
0 & (\frac{c_{2}}{c_{1}}-1) & 0 & 0 \\
0 & 0 & 0 & 0
\end{array}
\right).
\end{eqnarray}
Note that this matrix ${\Gamma}_{\phi}$ can be written in terms of the generator of rotation  $J_{12}$ about the z-axis
$
\label{eq4}
\Gamma_{\phi}= i\left(\frac{c_{2}}{c_{1}}-1\right)J_{12}.
$
We note that, if the velocities coincides $c_{1}=c_{2}$, the conical singularity in the origin is broken, henceforth the geometry of a disgyration is flat in all regions of  ``space-time". Let us analyze the parallel transport of vectors around closed curves that contains the singularity. The holonomy transformation (\ref{eq1}) for closed curves is defined by
\begin{equation}
\label{eq5}
U(\gamma)= {\cal P}\exp \left[ -\oint_{\gamma}dx^{\mu}
\Gamma_{\mu}(x) \right],
\end{equation}
where ${\cal P}$ is the order operator. From eq.(\ref{eq5}) we can construct the invariants which involve the holonomy group. This group measures the deviation of the space from global flatness. The basic invariant associated with the matrix $U(\gamma)$ is its trace. This quantity provides information about the geometric and physical structure of the space-time. In our case the unique contribution to the holonomy is provided by the azimuthal spin connection. Thus, we may write:
$U(\gamma)={\cal P}\exp\left(-\oint \Gamma_{\phi}d\phi \right)$. Making the expansion of this expression and noticing that we are always able to write the exponents of upper order in the $\Gamma_{\phi}$ and $\Gamma_{\phi}^{2}$ 
terms, the above equation can be written in the following way
\begin{eqnarray}
U(\gamma_{\phi}) =1-\frac{\Gamma_{\phi}}{c_{2}/c_{1}-1}\sin\left[
2\pi(c_{2}/c_{1}-1)\right]+ \frac{\Gamma_{\phi}^{2}}{(c_{2}/c_{1}-1)^{2}}
\left\{1-\cos\left([2\pi(c_{2}/c_{1}-1)\right)] \right\}
\end{eqnarray}
Alternatively one can express the holonomy in a matrix form
\begin{eqnarray}
\label{b10}
U(\gamma_{\phi})= \left(
\begin{array}{cccc}
1 & 0 & 0 & 0 \\
0 & \cos 2\pi(c_{2}/c_{1}-1) & \sin 2\pi(c_{2}/c_{1}-1) & 0 \\
0 & -\sin 2\pi(c_{2}/c_{1}-1) &\cos 2\pi(c_{2}/c_{1}-1) & 0  \\
0 & 0 & 0 & 1
\end{array}
\right).
\end{eqnarray}
This matrix can be interpreted as the rotation generator around the $z$-axis. It is well known that when $c_{2}=c_{1}$, the holonomy is trivial, But we expect that, when $c_{2}\neq c_{1}$, the holonomy would 
be always nontrivial and the deficit angle is given by $c_{2}/c_{1}-1$. But there exists certain values of the ratio $c_{2}/c_{1}$ in terms of the number of loops $n$ and an integer $m$ that makes the holonomy trivial~\cite{cqg:ellis}. This ratio is given by: $c_{2}/c_{1}=1+m/n.$ Therefore, from eq.(\ref{b10}) we conclude that when a vector is parallely transported along a closed curve surrounding the string, the transported vector acquires a non-zero phase factor. This non-trivial phase is an expression of the gravitational Aharonov-Bohm effect. This effect might be understood in terms of the global aspects of the underlying space-time manifold. In this way we obtain the phase acquired by quasiparticles in the background of a disgyration in a superfluid. The deficit angle $\chi$ obtained when we compare the final and initial position of the parallel transported vector is given by
$
\cos \chi_{A}=U_{A}^{A},
$
where $A$ is a tetradic index. The terms of non vanishing angular deviations occur when $A=1$ and $2$, so we have
$
\cos\chi_{1  or  2}=\cos(2\pi(c_{2}/c_{1}-1))$
or
$
|\chi_{1  or  2}|=|2\pi(c_{2}/c_{1}-1) + 2\pi n|.
$
For $2\pi(c_{2}/c_{1}-1) \rightarrow 0$ when $c_{2}=c_{1}$ we must have $\chi_{1 or 2} \rightarrow 0$,
so we choose $n=0$, which leads to
\begin{eqnarray}
\label{angu}
|\chi_{1  or  2}|=|2\pi(c_{2}/c_{1}-1)|.
\end{eqnarray}
The above expression shows that for $2\pi(c_{2}/c_{1}-1)\neq 0$, if we parallel transport a vector around a closed path, the final vector  does not coincide with the original vector. This physical effect could be understood as a gravitational analogue of the Aharonov-Bohm effect. We 
saw in (\ref{angu}) that there will be no Aharonov-Bohm effect if $2\pi(c_{2}/c_{1}-1)$ is an integer. This conditions is not always satisfied, because $2\pi(c_{2}/c_{1}-1)$ is not necessarily an integer. 
In fact, it can assume an arbitrary value. There are similarities of this effect with the Aharonov-Bohm effect in the cosmic string case~\cite{prd:b,valdir1}.  The crucial analogy occurs in the fact that the spacetime exterior to the cosmic string has null Riemann tensor. In this way when we transport the wave function, $\Psi$, around the defect we obtain the following result
$
\Psi'=U(\gamma)\Psi.
$
This relation provided the phase obtained by the wave function when it is transported around the defect. This result could be seen as the Aharonov-Bohm effect. This result can be used to describe the Aharonov-Bohm effect in other media with defects. For example, a class of defects studied by Volovik: vortex with fractional quantum number,
$ N=1/2$ in chiral superfluid and $N=1/2$ and $N=1/4$ in chiral superconductors. In $^{3}He-A$  the vortex with half integer 
$N=2m/h \oint d \mathbf{r} \cdot \mathbf{v}_{s}$ results of the combination of the $\pi$-disclination in $\mathbf{d}$ and a $\pi$-vortex. This defect is characterized with a condensed matter analog of the Alice string~\cite{volouf:jetp} in particle physics. The quasiparticles going around a $1/2$ vortex flips its $U(1)_{s_{3}}$ charge, its spin.  This fact is responsible by global the Aharonov-Bohm~\cite{jetp:kha,prl:pres} effect in $^{3}He-A$. The contribution of global Aharonov-Bohm effect of a $\pi$-disclination is given by $U(\gamma)$ and the global Aharonov-Bohm effect is given by a global phase factor. We analyze now the defect in a superconductor that combines the fractional flux trapped by loops of monocrystal 
 with tetragonal symmetry. In this   case the deformation contributes to the superfluid current and thus to the angular momentum of the source, this contribution has been added to the spin connection. The defect is represented by a $\pi/2$  disclination and the flux is contained inside the disclination core. We use here the same notation due to Volovik~\cite{volo:pas} that  the flux carried by a conventional $N=1$ abrikosov vortex in a conventional superconductor is $\frac{1}{2}\Phi_{0}$ where $\Phi_{0}=hc/e$. 
Our approach used in the disgyration case can be used to describe the parallel transport of a vector in the background of a disclination in a superconductor lattice. The metric that describes the disclination in a crystal is similar to the disgyration metric were the parameter $a=c_{2}/c_{1}$ in disgyration metric is substituted by $a=1\pm \lambda/2\pi$, were $\lambda$ is the angular sector removed or inserted by creation of the disclination. In this way the holonomy due to the disclination is equal to expression (\ref{b10}) were we substituted the parameter $a$ of the disgyration by the correspondent parameter of the disclination. But this defect is a combination of the vortex with a disclination. The vortex contribution to holonomy is due to a vector potential $\mathbf{A}=(mc/e)\mathbf{v}_{s}$ and to a magnetic flux
$\int d \mathbf{S} \cdot \mathbf{B}=\oint d \mathbf{r} \cdot \mathbf{A}=(mc/e) \oint d \mathbf{r} \cdot \mathbf{v}_{s} = (N/2)\Phi_{0}$. 
In this way we introduce a global contribution to the holonomy due to the vector potential given by a global Aharonov-Bohm phase 
$e^{i\oint A(r) \cdot dr}$~\cite{volo:pas}. 
A new contribution can be included in this holonomy due to a internal flux in the defect that represents the vorticity of the defect. In this way
\begin{eqnarray}
\Psi'= e^{i\oint A(r) \cdot dr}U(\gamma)\Psi,
\end{eqnarray} 
where $ \mathbf{A}=(mc/e)\mathbf{v}_{s}$ for a supeconductor. For the case of a $\pi/2$-disclination the matrix of parallel transport is given after $n$ loops around of superconductor vortex is given by
\begin{eqnarray}
\label{b110}
U(\gamma_{\phi})= \left(
\begin{array}{cccc}
1 & 0 & 0 & 0 \\
0 & \cos \frac{n \pi}{2} & -\sin \frac{n \pi}{2} & 0 \\
0 & \sin \frac{n \pi}{2} &\cos \frac{n \pi}{2} & 0  \\
0 & 0 & 0 & 1
\end{array}
\right).
\end{eqnarray}
The matrix (\ref{b110}) gives  the phase that a vector has when transported in a loop around of disclination. This is a geometric contribution to the Aharonov-Bohm effect. The vortex with $N=1/2$ in a superconductor has been observed in a high-temperature superconductor~\cite{prl:kirtley}. In this case we have a contribution of the vector potential and a geometric contribution of the $\pi/2$-disclination. For a fractional vortex with $N=1/2$, in a superfluid, we have two contribution: one due to the global    Aharonov-Bohm effect\cite{jetp:kha,prl:pres} and the other to a global  Aharonov-Bohm effects due the geometrical aspects of crystal deformation caused by $\pi$-disclination.


In this letter we investigate a quasiparticle in the presence of disgyration by making use of holonomy transformations. Analyzing equation (\ref{b10}) we note that, for $c_{1}=c_{2}$, the holonomy is trivial, but otherwise the holonomy matrix is not trivial, characterizing the gravitational analogue of the Aharonov-Bohm effect for phonons in a superfluid. The conical nature of the background ``space-time" that describes the disgyration is responsible by the ``gravitational" Aharonov-Bohm for quasiparticle in this system. This fact can be used to investigate this phenomenon in laboratory.  This type of Aharonov-Bohm effect is characterized by: in the external region of the defect where the curvature is null, a characteristic of conical singularities, the holonomy demonstrates that the vectors acquires a angle when parallely transported around the defect. Note that in the present case the nontrivial ``space-time" structure, which is locally flat but globally not equivalent to a Minkowski space-time, is responsible by the non-local aspect of the gravitational analogue of the Aharonov-Bohm effect. Notice that this effect is similar to the effect experimented by particles in the background of a cosmic string, well known as the gravitational Aharonov-Bohm effect~\cite{Ford,valdir1}. In contrast with the vortex case where, the analogy with the spinning cosmic string~\cite{prl:maz} demonstrated that the angular momentum is responsible by the gravitational Aharonov-Bohm effect~\cite{boo,fisher1}. In the disgyration case the Aharonov-Bohm effect is a global effect due to topological properties of the conical space-time. We claim that this formalism can be used to analyze the Aharonov-Bohm effect in a chiral superconductor and a chiral superfluid. We introduce the holonomy of these defects considering the contribution of the gauge field and the geometric contribution of        disclinations. The contribution to the holonomy given by the internal flux was introduced {\it ad-hoc}, considering that this contribution is well known in the literature. In contrat with this defect in the disgyration case, ``electromagnetic" effects of the l-field is absent, and one a has pure gravitational field~\cite{jetp:volomono}. This happens if the ``magnetic"($\nabla \times l=0$) field is absent. In this way the spin connection does not give any contribution to this magnetic field and consequently to the Aharonov-Bohm effect. This approach can also be used to analyze the geometric phase for phonons in the presence of a vortex in a superfluid. Examples of this type demonstrate the power of condensed matter systems as analog models for general relativity.
  
We thank to CAPES (PROCAD), FAPESQ-PB, CNPq and PRONEX for
financial support. We thank Professor F. Moraes for the critical reading of this manuscript. One of us (GdA) thanks Professors Volovik, Leonhardt, Kleinert and Mazur for helpful discussions on the subject of this paper. 


\begin{thebibliography}{10}
\bibitem{prl:unruh}W. G. Unruh, Phys. Rev. {\bf D 51}, 2827 (1995);ibid.,Phys. Rev. D, {\bf 51} 6(1995).
\bibitem{prl:gar} L. J. Garay, J. R. Anglin, J. I. Cirac, and P. Zoller
Phys. Rev. lett. {\bf 85}, 4643 (2000);  L. J. Garay, J. R. Anglin, J. I. Cirac and P. Zoller Phys.
Rev. {\bf 63} 023611 (2001).
\bibitem{prd:jac} T. Jacobson, Phys. Rev. {\bf D 44}, 1731 (1991); M. Visser, Phys. Rev. Lett. {\bf 80}, 3436 (1998); Matt Visser Class. Quant. Grav. {\bf 15} 1767 (1998).
\bibitem{volo} G. E. Volovik, JETP LETT, {\bf 67} (11) (1998).
\bibitem{boo} G. E. Volovik {\it The Universe in a Helium Droplet}( Oxford
University Press)2003.
\bibitem{leon} Ulf Leonhardt, Phys. Rev. {\bf A 65}, 043818 (2002); I. Brevik et al., Phys. Rev. {\bf D 65}, 024005 (2002).
\bibitem{prd:novello} M. Novello, Int. J. Mod. Phys A {\bf 17} 29, 4187
(2002).
\bibitem{3} G. E. Volovik, Low Temp. Phys. {\bf 24}, 2 (1998).
\bibitem{staro}D. D. Sokolov and A. A. Starobinskii, Sov. Phy. Dokl {\bf 22} 312 (1977) 
\bibitem{katanaev} M. Katanaev and I. Volovich, Annals of Physics (N.Y.) {\bf 216}, 1(1992); E. Kr\"{o}ner,Continuum Theory of Defects,Les
Houches-Physics of Defects (1980). H. Kleinert, \textit{Gauge fields in
Condensed Matter},
Vols. I, II (World Scientific, Singapore, 1989).
\bibitem{1} C. Furtado and F. Moraes, Europhysics Letters {\bf 45} 279 (1999).
\bibitem{Vilenkin} Vilenkin A., 1981 Phys. Rev. {\bf D23}, 852; Linet B., 1985 Gen. Rel. Grav. {\bf 17}, 
1109.
\bibitem{Ford} Ford L. H. and Vilenkin A., 1981 J. Phys. {\bf A14}, 2353; Dowker J. S., 1967 Nuovo Cimento 
{\bf 52}, 129; Anandan J., 1994 Phys. Lett. {\bf  A195}, 284.

\bibitem{pr:aha} Aharonov Y. and Bohm D., 1959 Phys. Rev. {\bf 115}, 485.

\bibitem{Frolov} Frolov V. P., Skarzhinsky V. D. and John R. W., 1987 II Nuovo Cim., {\bf 99B}, 67.
\bibitem{prl:maz}P.O. Mazur Phys. Rev. Lett. {\bf 57} 929 (1986);P.O. Mazur Phys. Rev. Lett. {\bf 59} 2380 (1987).
\bibitem{prd:b}Burgers, C J C. 1985 Phys. Rev. D, 32 (02)
\bibitem{valdir1} V. B. Bezerra, Phys. Rev. {\bf D35}, 2031 (1987); V. B. Bezerra, Ann. Phys. (NY) {\bf 203} 392 (1990)
\bibitem{law}  J. K. Lawrence, D. Leiter and G. Szamosi, Nuovo Cimento {\bf
17B}, 113 (1973);  V. P. Frolov and V. D. Skarzhinsky, Nuovo Cimento  {\bf 99B},
67 (1987).
\bibitem{man} Mandelstam S, Ann. Phys.(N.Y.), {\bf19} 1 and 25 (1962).
\bibitem{wu} Wu T. T. and Yang C. N. Phys. Rev. {\bf D12} 3845 (1975);
{\bf D14} 437 (1976).
\bibitem{vor} Voronov N A and Makeenko Y M ,Sov. J. Nuc. Phys. {\bf 36} 444
(1982).
\bibitem{cqg:ellis}T. Rothman, G. Ellis and J. Murugan.
Class. Quant. Grav. {\bf 18} 1217, (2001) .
\bibitem{volouf:jetp} U. Leonhardt and G. E. Volovik JETP Lett. {\bf 72} 46 (2000).
\bibitem{jetp:kha}M. V. Khazan, JETP Lett. {\bf 41} 486 (1985).
\bibitem{prl:pres}J. March-Russel, J. Preskill and F. Wilczek, Phys Rev. Lett. {\bf 50} 2567 (1992); A. C. Davis and P. Martim, Nucl. Phys. {\bf B 419} 341 (1994).
\bibitem{volo:pas} G. E. Volovik Proc.Nat.Acad.Sci. {\bf 97} 2431 (2000) 
\bibitem{prl:kirtley} J. R. Kirtley, C. C. Tsuei, M. Rupp et al Phys. Rev. Lett. {\bf 76} 1336 (1996).
\bibitem{fisher1} U. R. Fischer  and M. Visser, Phys. Rev. Lett. {\bf 88}, 110201 (2002);U. R. Fischer and M. Visser , Ann. Phys. (NY) {\bf 304} 22 (2003);Michael Stone, Phys Rev , {\bf B 61} 11780  (2000).
\bibitem{jetp:volomono} G. E. Volovik JETP Lett. {\bf 67} 698 (1998).
\end{thebibliography}
\end{document}